\documentclass[prl,showpacs,superscriptaddress,showkeys]{revtex4}
\usepackage{graphicx}
\usepackage{dcolumn}
\usepackage{bm}

\begin{document}

\title{Ultrahigh Light Intensification \\
by a Counter-Propagating Breaking Plasma Wave \\
-- Relativistic Flying Parabolic Mirror%
\thispagestyle{headings}}

\author{Sergei V. Bulanov}
 \altaffiliation[Also at ]{General Physics Institute RAS, Vavilov str. 38, Moscow 119991, Russia}

\author{Timur Esirkepov}
 \altaffiliation[Also at ]{Moscow Institute of Physics and Technology, Institutskij per. 9, Dolgoprudny Moscow region 141700, Russia}
 \email{timur@apr.jaeri.go.jp}

\author{Toshiki Tajima}
\affiliation{Kansai Research Establishment, JAERI, Kizu, Kyoto 619-0215, Japan}

\date{May, 2003}

\begin{abstract}
A method to generate ultrahigh intense electromagnetic fields is suggested,
based on the laser pulse compression, carrier frequency
upshift and focusing
by a counter-propagating breaking plasma wave,
relativistic flying parabolic mirror.
This method allows us to achieve the quantum electrodynamics critical field
(Schwinger limit)
with present day laser systems.
\end{abstract}

\pacs{12.20.-m, 13.40.-f, 52.27.Ny, 52.38.-r, 52.65.Rr}

\keywords{quantum electrodynamics critical field, relativistic plasmas,
ultrashort laser pulse, laser-plasma interaction, wakefield,
particle-in-cell simulation}

\maketitle


The invention of chirped pulse amplification (CPA) method and recent
development of laser technology led to a stunning increase of the light
intensity in a laser focal spot, \cite{Mourou2}.
Electrons in laser electromagnetic
field become relativistic at intensities $I\sim 10^{18}$W/cm$^{2}$. The ion
motion strongly affects the relativistic plasma dynamics starting from $%
I\gtrsim (m_{i}/m_{e})\times 10^{18}$W/cm$^{2}$ (see Ref. \cite{B-review}
and references therein). Nowadays lasers produce pulses whose intensity is
approaching to $10^{22}$W/cm$^{2}$ \cite{Mourou2}. With further increase
of intensity we shall meet novel physical processes such as the radiation
reaction dominated regimes, which come into
play at $I=10^{23}-10^{24}$W/cm$^{2}$ \cite{Zhidkov}, and
then the regime beyond $I=10^{25}$W/cm$^{2}$
where the quantum electrodynamics (QED) description
is needed as
the recoil of emitted photon momentum becomes comparable
with the electron momentum \cite{B2}.
Near the intensity of $10^{29}$W/cm$^{2}$,
corresponding to the QED critical electric field, light can
generate electron-positron pairs from vacuum \cite{el-pos1, el-pos2}.
Even before that limit, vacuum begins to act nonlinearly such as vacuum polarization.
These nonlinear effect have attracted a great deal of attention since
\cite{el-pos1}: they lie outside the scope of perturbation theory and
shed  light on the nonlinear quantum
electrodynamics properties of the vacuum. There are several ways to achieve
such an intensity. One way was demonstrated in the experiments \cite{SLAC96}%
, where high-energy bunch of electrons interacts with counterpropagating
intense laser pulse. In the reference frame of electrons the electric field
magnitude of the incident radiation was approximately 25\% of the QED
critical field.

A technically feasible way is to increase the power of the
contemporary laser system by some 7 orders of magnitude
through megajoule lasers \cite{TM}, albeit quite expensive.
Another way is
to increase the frequency of the laser radiation and then focus it onto a
tiny region. In this method X-ray lasers can be used \cite{Ringwald}. To
achieve more ``moderate'' intensities, $10^{24}-10^{25}$W/cm$^{2}$, another
scheme was suggested in Ref. \cite{Shen-Yu02}, where a quasi-soliton wave
between two foils is pumped by the external laser field up to an ultrahigh
magnitude.
Another method is based on the simultaneous laser frequency
upshifting and the pulse compression.
These two phenomena were demonstrated in a broad variety of configurations,
where they were caused, in general, by different mechanisms.
In particular,
the wave amplification reflected at the moving relativistic electron slab
was discussed in Refs. \cite{Landecker-52}
(based on the frequency up-shift of radiation reflected at the relativistic
mirror, as predicted by A. Einstein in Ref. \cite{Ein});
the backward Thompson scattering at relativistic electron bunch
was considered in Refs. \cite{Arutyunian-63};
the reflection at the moving ionization fronts has
been studied in Refs. \cite{Semenova-67};
``photon acceleration''schemes with co-propagating laser pulses
in underdense plasma were examined in Refs. \cite{Photon-Accelerator};
various schemes of the
counter-propagating laser pulses and the use of
parametric amplification process were discussed in Refs. \cite{Shvets-98}.


In the present paper we consider a plasma wakefield in the
{\it  wave-breaking regime} as a tool for generating a coherent radiation of
ultra-high intensity. Compared to the previously discussed schemes this
regime demonstrates both the robustness and coherence of the transformed
laser light.

We examine the following scenario. A short intense laser pulse (the``%
{\it  driver} pulse'') induces wakefield in a plasma.
As it is well known \cite{Tajima-Dawson},
the wakefield phase velocity $v_{ph}=\beta _{ph}c$ equals the laser pulse
group velocity, which is close to the speed of light in a vacuum when the
laser pulse propagates in the underdense plasma. The corresponding Lorentz
factor is $\gamma _{ph}=(1-\beta _{ph}^{2})^{-1/2}\approx \omega _{d}/\omega
_{pe}$, where $\omega _{d}$ is the {\it  driver} pulse frequency, $\omega
_{pe}$ is the Langmuir frequency. The nonlinearity of strong wakefield
causes a nonlinear wave profile, including a steepening of the wave and
formation of localized maximums in the electron density -- the spikes \cite
{Akhiezer-Polovin}. This amounts to wavebreaking regime (see Ref. \cite
{B-review} and references therein). Theoretically the electron density in
the spike tends to infinity, but remains integrable \cite{B-review}.
Sufficiently weak counter-propagating laser pulse (the ``{\it  source}
pulse'') will be partially reflected from the density maximum.
The
reflection coefficient scales as $\gamma _{ph}$
and the reflected wave vector-potential scales as $\gamma _{ph}^{-3}$,
as it is shown below.
As we
see, the electron density maximum acts as a mirror flying with the
relativistic velocity $v_{ph}\approx c$. The frequency of the reflected
radiation is up-shifted by factor $(1+\beta _{ph})/(1-\beta _{ph})\approx
4\gamma _{ph}^{2}$, in accordance with the Einstein formula \cite{Ein}.
It
is important that the relativistic dependence of the Langmuir frequency on
the {\it  driver} pulse amplitude causes parabolic bending of
constant phase surface of the plasma wave, since the {\it  driver}
pulse has a finite
transverse size \cite{D-shape}.
As a result, the surface where the
electron density is maximal has a shape close to a paraboloid.
Because we have a curved mirror, the frequency $\tilde{\omega}_{s}$ of the
reflected radiation depends on the angle: 
\begin{equation}
\tilde{\omega}_{s}=\frac{1+\beta _{ph}}{1-\beta _{ph}\cos \theta }\omega
_{s}\,,  \label{eq:freq}
\end{equation}
where $\omega _{s}$ is the {\it  source} pulse frequency, and $\theta $ is
the angle between the reflected wave vector and the direction of the driver
pulse propagation in the laboratory frame. The curved mirror focuses the
reflected light. The focal spot size is of the order of the diffraction
limited size. In the reference frame of the wakefield it is $\lambda
_{s}^{\prime }=\lambda _{s}((1-\beta _{ph})/(1+\beta _{ph}))^{1/2}\approx
\lambda _{s}/2\gamma _{ph}$, where $\lambda _{s}$ is the wavelength of the 
{\it  source} pulse. In the laboratory frame the focal spot size is
approximately $\lambda _{s}/4\gamma _{ph}^{2}$ along the paraboloid axis,
and $\approx \lambda _{s}/2\gamma _{ph}$ in the transverse direction. In the
focal spot the resulting intensity gain factor scales with $\gamma _{ph}$ as 
$\gamma _{ph}^{-3}\times (\tilde{\omega}_{s}/\omega _{s})^{2}\times
(D_{s}/\lambda _{s}^{\prime })^{2}=64(D_{s}/\lambda _{s})^{2}\gamma _{ph}^{3}
$, where $D_{s}$ is the diameter of the efficiently reflected portion of the 
{\it  source} pulse beam. This value can be great enough to substantially
increase the intensity of the reflected light in the focus, even up to the
QED critical electric field.


In order to calculate the reflection coefficient, we consider the
interaction of an electromagnetic wave with a maximum of the electron
density formed in a breaking Langmuir wave. In the laboratory frame, this
interaction can be described by the wave equation 
\begin{equation}
\partial _{tt}A_{z}-c^{2}\Delta A_{z}+\frac{4\pi e^{2}n\left(
x-v_{ph}t\right) }{m_{e}\gamma _{e}}A_{z}=0\,,  \label{eq-Wave-lab}
\end{equation}
where $A_{z}$ is the $z$-component of the vector potential, $\gamma _{e}$ is
the electron Lorentz factor, and $\gamma _{e}\approx \gamma _{ph}$ near the
maximum of the density in the wakewave wavebreaking regime.

According to the continuity equation $\partial _{t}n_{e}+{\rm   div}%
(n_{e}v_{e})=0$, the electron density in the stationary Langmuir wave is
given by $n=n_{0}v_{ph}/(v_{ph}-v_{e})$, where the electron velocity $v_{e}$
varies from $-v_{ph}$ to $v_{ph}$ (see Ref. \cite{Akhiezer-Polovin}), and
the electron density varies from the minimal value $=n_{0}/2$ to infinity
(integrable). For the breaking plasma wakewave, in every wave period
approximately a half of electrons are located in the spike of the electron
density. Therefore we can approximate the electron density by $n\left(
x-v_{ph}t\right) =(1+\lambda _{p}\delta (x-v_{ph}t))n_{0}/2$, where $\lambda
_{p}$ is the wakefield wavelength and $\delta (x)$ is the Dirac delta
function. This approximation is valid when the density maximum thickness is
sufficiently less than the collisionless skin depth $c/\omega _{pe}$ and 
{\it  source} pulse wavelength in the wakefield rest frame, i. e. when the
wakefield is close to the wave-breaking regime.

In the reference frame comoving with the plasma wakewave, Eq. (\ref
{eq-Wave-lab}) has the same form. The Lorentz transformation to this
frame is given by $t^{\prime }=(t-v_{ph}x/c^{2})\gamma _{ph}$, $%
x^{\prime }=(x-v_{ph}t)\gamma _{ph}$, $y^{\prime }=y$, $z^{\prime }=z$.

We seek for solution to Eq. (\ref{eq-Wave-lab}) 
in the form 
$A_{z}={\cal   A}(x^{\prime })\exp \left( i(\omega _{s}^{\prime }t^{\prime }+k_{x}^{\prime
}x^{\prime }+k_{y}^{\prime }y^{\prime }+k_{z}^{\prime }z^{\prime })\right) $, 
where 
$\omega _{s}^{\prime }=(\omega _{s}+v_{ph}k_{x})\gamma _{ph}$, $k_{x}^{\prime }=(k_{x}+v_{ph}\omega /c^{2})\gamma _{ph}$, $k_{\perp
}^{\prime }=k_{\perp }$ are the frequency and wavevector in the moving
frame, and $k_{x}^{\prime }>0$. Using this ansatz, from Eq. (\ref
{eq-Wave-lab}) in the moving frame we obtain 
\begin{equation}
\frac{d^{2}{\cal   A}}{dx^{\prime }{}^{2}}+q^{2}{\cal   A}=\chi \delta
(x^{\prime }){\cal   A}\,,  \label{eq:A}
\end{equation}
where $q^{2}=\omega _{s}^{\prime }{}^{2}/c^{2}-k_{\perp }^{\prime
}{}^{2}-\omega _{pe}^{2}/(2c^{2}\gamma _{ph})>0$ and $\chi =\omega
_{pe}^{2}\lambda _{p}/c^{2}$. This equation is equivalent to the scattering
problem at the delta potential. The solution is ${\cal   A}(x^{\prime
})=\exp (iqx^{\prime })+\rho (q)\exp (-iqx^{\prime })$ for $x^{\prime }\ge 0$
(incident and reflected wave), and ${\cal   A}(x^{\prime })=\tau (q)\exp
(iqx^{\prime })$ for $x^{\prime }<0$ (transmitted wave), where $\rho
(q)=-\chi /(\chi +2iq)$ and $\tau (q)=iq/(\chi +2iq)$. In a nonlinear
Langmuir wave, its wavelength depends on the wave amplitude \cite
{Akhiezer-Polovin}, and for the breaking wakewave we have $\lambda
_{p}\approx 4(2\gamma _{ph})^{1/2}c/\omega _{pe}$. In this case $\chi
=4(2\gamma _{ph})^{1/2}\omega _{pe}/c$. Taking $\omega
_{s}^{\prime }{}^{2}=4\gamma _{ph}^{2}\omega _{s}^{2}$
into account, we find that the
reflection coefficient, defined as a ratio of the reflected to the
incident energy flux,  in the co-moving frame is $\approx \left( \omega
_{d}/\omega _{s}\right) ^{2}/2\gamma _{ph}^{3}$. In the laboratory frame it
is 
\begin{equation}
R\approx 8\gamma _{ph}\left( \omega _{d}/\omega _{s}\right) ^{2}\,.
\label{eq:refl}
\end{equation}

The intensity $\tilde{I}_{sf}$ in the focal spot of the {\it  source}
pulse, reflected and focused by the electron density maximum in the
laboratory frame, is increased by the factor of the order of 
\begin{equation}
\tilde{I}_{sf}/I_{s}\approx 32(\omega _{d}/\omega _{s})^{2}(D_{s}/\lambda
_{s})^{2}\gamma _{ph}^{3}\,.  \label{eq:factor}
\end{equation}
Theoretically, the actual gain can be even greater, because a) the estimation (%
\ref{eq:refl}) corresponds to one-dimensinal case, whereas the density
modulation in the 3D breaking wakewave is stronger, b) the reflectance (\ref
{eq:refl}) of the 3D paraboloidal mirror is greater at the periphery.


We consider the following example. A one-micron laser pulse ({\it  driver}%
) generates wakefield in a plasma with density $n_{e}=10^{17}$cm$^{-3}$. The
corresponding plasma wavelength is $\lambda _{p}\approx 100\mu $m. The
Lorentz factor, associated with the phase velocity of the wakefield, is
estimated as $\gamma _{ph}\approx \omega _{d}/\omega _{pe}\approx 100$. The
counter-propagating one-micron laser pulse with intensity $I_{s}=10^{17}$W/cm%
$^{2}$ ({\it  source}) is partially reflected and focused by the wakefield
cusp. If the efficiently reflected beam diameter is $D_{s}=200\mu $m, then,
according to Eq. (\ref{eq:factor}), the final intensity in the focal spot is 
$\tilde{I}_{sf}\approx 1.5\times 10^{29}$W/cm$^{2}$. The {\it  driver}
pulse intensity should be sufficiently high and its beam diameter should be
enough to give such a wide mirror, assume $I_{d}=10^{18}$W/cm$^{2}$ and $%
D_{d}=800\mu $m. Thus, if both the {\it  driver} and {\it  source} are
one-wavelength pulses, they carry  $17$J and $0.1$J, respectively. We see
that in an optimistic scenario the QED critical electric field may be
achieved with the present-day laser technology!


\begin{figure}
\includegraphics{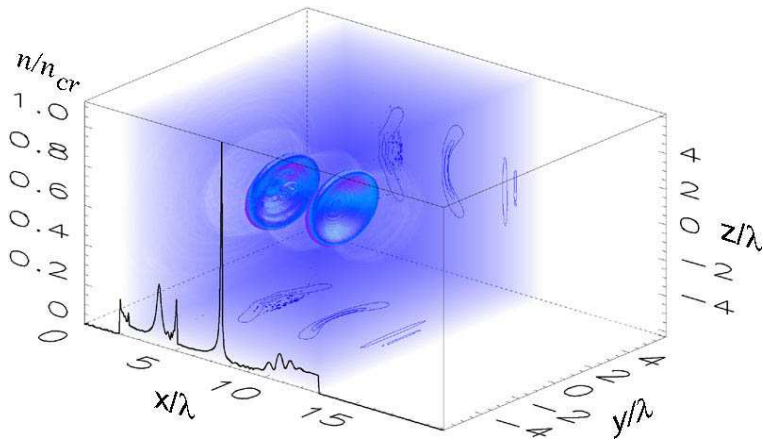}
\caption{The electron density in the wake of the {\it  driver} laser pulse
at $t=14\times 2\pi/\omega_d$.
The $(x,y=-6\lambda,z)$-plane: density profile along the symmetry axis.
Blue curves for density values $n=0.12,0.24,0.36 n_{cr}$ on the
corresponding perpendicular planes of symmetry.
Isosurfaces for value $n=0.15 n_{cr}$, ``blue gas'' for lower values.
}
\label{fig:wake}
\end{figure}

\begin{figure}
\includegraphics{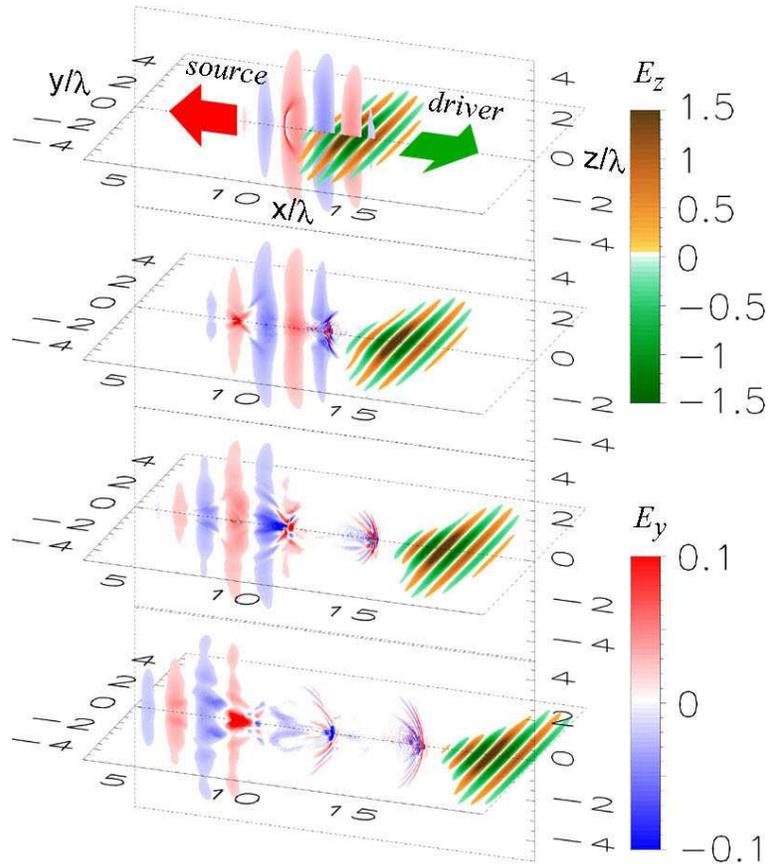}
\caption{The cross-sections of the electric field components.
The ($x,y,z=0$)-plane: $E_z(x,y,z=0)$ (green-brown colorscale),
the plane ($x,y=0,z$): $E_y(x,y=0,z)$
(blue-red colorscale) at
$t=16,18,20,22\times 2\pi/\omega_d$ (top-down). }
\label{fig:E}
\end{figure}

To demonstrate the feasibility of the effect of the light reflection and
focusing by the breaking wakewave, we performed three-dimensional
particle-in-cell (PIC) simulations using the code REMP (Relativistic
Electro-Magnetic Particle-mesh code) based on scheme \cite{E-CPC}. In the
simulations the {\it  driver} pulse propagates in the direction of the $x$%
-axis. Its dimensionless amplitude is $a_d=1.7$ which corresponds to peak
intensity $4\times 10^{18}$W/cm$^2 \times (1\mu{\rm   m}/\lambda_d)^2$,
where $\lambda_d$ is the {\it  driver} wavelength. The {\it  driver} is
linearly polarized along the $z$-axis, it has the gaussian shape, its FWHM size
is $3\lambda_d \times 6\lambda_d \times 6\lambda_d$. The {\it  source}
pulse propagates in the opposite direction. Its wavelength is two times
greater than the {\it  driver} wavelength, $\lambda_s =
2\lambda_d$. The {\it  source} pulse amplitude is chosen to be small, $%
a_s=0.05$, to reduce the distortion of the wakewave. The pulse shape
is rectangular in the $x$-direction and Gaussian in the transverse
direction, its size is $6\lambda_d \times 6\lambda_d \times 6\lambda_d$. To
distinguish the electromagnetic radiation of the {\it  driver} from the
{\it source} pulses, we set the {\it  source} pulse to be linearly polarized in
the direction perpendicular to the {\it  driver} polarization, i. e.
along the $y$-axis. The laser pulses propagate in the underdense plasma slab
with the electron density $n_e=0.09n_{cr}$, which corresponds to the
Langmuir frequency $\omega_{pe}=0.3\omega_d$. The plasma slab is localized
at $2\lambda_d<x<13\lambda_d$ in the simulation box with size $22\lambda_d
\times 19.5\lambda_d \times 19.2\lambda_d$. The simulations were carried out
on 720 processors of the supercomputer HP Alpha Server SC ES40 at
JAERI Kansai.
The mesh size is $dx=\lambda_d/100$, total number of quasiparticles is $10^{10}$
(ten billion). The boundary conditions are absorbing on the $x$-axis and
periodic in the transverse directon, both for the electromagnetic fields and
quasi-particles.
We emphasize that the simulation grid
must be and in fact was chosen to be
fine enough to resolve the huge frequency up-shift given by Eq.(\ref{eq:freq}),
exhausting all the supercomputer resources.

The simulation results are presented in Figs. \ref{fig:wake} and \ref{fig:E}.
Fig. \ref{fig:wake} shows the plasma wakewave induced by the {\it  driver}
laser pulse as modulations in the electron density. We see the electron
density cusps in the form of paraboloids. They move with velocity $%
v_{ph}\approx 0.87c$, the corresponding gamma-factor is $\gamma_{ph}\approx
2 $. Their transverse size is much larger than the wavelength of the
counterpropagating {\it  source} pulse in the reference frame of the
wakefield.
As seen from the electron density profile along the axis
of the {\it  driver} pulse propagation,
the wakewave dynamics is
close to wave-beaking regime. Each electron density maximum forms a
semi-transparent parabolic mirror, which reflects a part of the
{\it  source} pulse radiation.

In Fig. \ref{fig:E} we present the electric field components.
The {\it  driver} pulse is seen in the
cross-section of the $z$-component of the electric field in the ($x,y,z=0$%
)-plane. The {\it  source} pulse and its reflection is seen in the
cross-section of the $y$-component of the electric field in the ($x,y=0,z$%
)-plane. The part of the {\it  source} pulse radiation is reflected from
the flying paraboloidal mirrors, then it focuses yielding the peak intensity
in the focal spot, and finally it defocuses and propagates as a spherical
short wave train, whose frequency depend on the wave vector direction, in
agreement with Eq.(\ref{eq:freq}).
This process is clearly seen in the
animations produced from the data (see authors' website).
The main part of the reflected light
power is concentrated whithin the angle $\sim 1/\gamma _{ph}$, hence this
coherent high-frequency beam resembles a searchlight. The reflected part has
the same number of cycles as the {\it  source} pulse, as expected, since
it is Lorentz invriant. The wavelength and duration of the reflected
pulse are approximately 14 times less than the wavelength and duration of
the {\it  source} pulse, in agreement with Eq.(\ref{eq:freq}) since $%
(1+\beta _{ph})/(1-\beta _{ph})\approx 14.4$. The focal spot size of the
reflected radiation is much smaller than the wavelength of the {\it  source%
} pulse. The electric field in the focal spot is approximately 16 times
higher than in the {\it  source} pulse. Therefore, the intensity increases
256 times in agreement with estimation (\ref{eq:factor}).

We emphasize that the efficient reflection is achievable only when the
wakefield is close to the wave-breaking regime and the cusps in the
electron density are formed. As we see in the simulations, the reflection
and focusing is robust and even distorted (to some extent) wakewave
can efficiently reflect and focus the {\it  source} pulse radiation. We
also observe that despite the moderate reflection coefficient, the colossal
frequency up-shift and focusing by a sufficiently wide (transversely)
wakewave give us a huge increase of the light intensity.

Similar processes may occur in laser-plasma interation spontaneously, e.g.
when a short laser pulse exciting plasma wakewave is a subject of the
stimulated backward Raman scattering or a portion of the pulse is
reflected back from the plasma inhomogeneity. Then the backward scattered
electromagnetic wave interacts with plasma density modulations in the
wakewave moving with relativistic velocity. According to scenario described
above, the electromagnetic radiation, reflected by the wakewave, propagates
in the forward direction as a high-frequency strongly collimated
(within the angle $\sim 1/\gamma _{ph}$) electromagnetic beam.


We have proposed the scheme of the relativistic plasma wake caustic light
intensification, which can be achieved due to the reflection and focusing of
light from the maximum of the electron density in the plasma wakewave
at close to the wave-breaking regime. The presented results of 3D PIC
simulations provide us a proof of principle of the electromagnetic field
intensification during reflection of the laser radiation at the flying
paraboloidal relativistic mirrors in the plasma wakewave. With the ideal
realization of the described scheme we can achieve extremely high
electric fields (in the laboratory reference frame) aproaching the QED
critical field with the present-day laser technology.
We envision the present example is just one manifestation
of what we foresee
as the emergence of {\it relativistic engineering}.

\begin{acknowledgments}
We appreciate the help of APRC computer group.
We thank V. N. Bayer, M. Borghesi, J. Koga, K. Mima, G. Mourou, N. B. Narozhny, 
K. Nishihara, V. S. Popov, A. Ringwald, V. I. Ritus, 
V. I. Telnov, and M. Yamagiwa for discussion.
\end{acknowledgments}


\end{document}